\UseRawInputEncoding
\documentclass[conference]{IEEEtran}
\IEEEoverridecommandlockouts
\usepackage{cite}
\usepackage{amsmath,amssymb,amsfonts}
\usepackage{algorithmic}
\usepackage{graphicx}
\usepackage{textcomp}
\usepackage{xcolor}
\def\BibTeX{{\rm B\kern-.05em{\sc i\kern-.025em b}\kern-.08em
    T\kern-.1667em\lower.7ex\hbox{E}\kern-.125emX}}

\makeatletter
\renewcommand{\maketag@@@}[1]{\hbox{\m@th\normalsize\normalfont#1}}%
\makeatother
\begin{document}

%
\title{RFC-HyPGCN: A Runtime Sparse Feature Compress Accelerator for Skeleton-Based GCNs Action Recognition Model with Hybrid Pruning
}
\author{\IEEEauthorblockN{Dong Wen, Jingfei Jiang, Jinwei Xu, Kang Wang, Tao Xiao, Yang Zhao, Yong Dou}
\IEEEauthorblockA{\textit{National University of Defense Technology}, Changsha, China
\\
{\{wendong19, jingfeijiang, xujinwei13, wangkang, xiaotao19, zhaoyang10, yongdou\}}@nudt.edu.cn}
}

\maketitle

\begin{abstract}

Skeleton-based Graph Convolutional Networks (GCNs) models for action recognition have achieved excellent prediction accuracy in the field. However, limited by large model and computation complexity, GCNs for action recognition like 2s-AGCN have insufficient power-efficiency and throughput on GPU. Thus, the demand of model reduction and hardware acceleration for low-power GCNs action recognition application becomes continuously higher.

To address challenges above, this paper proposes a runtime sparse feature compress accelerator with hybrid pruning method: RFC-HyPGCN. First, this method skips both graph and spatial convolution workloads by reorganizing the multiplication order. Following spatial convolution¡¯s channel-pruning dataflow, a coarse-grained pruning method on temporal filters is designed, together with sampling-like fine-grained pruning on time dimension. Later, we come up with an architecture where all convolutional layers are mapped on chip to pursue high throughput. To further reduce storage resource utilization, online sparse feature compress format is put forward. Features are divided and encoded into several banks according to presented format, then bank storage is split into depth-variable mini-banks. Furthermore, this work applies quantization, input-skipping and intra-PE dynamic data scheduling to accelerate the model. In experiments, proposed pruning method is conducted on 2s-AGCN, acquiring 3.0x-8.4x model compression ratio and 73.20\% graph-skipping efficiency with balancing weight pruning. Implemented on Xilinx XCKU-115 FPGA, the proposed architecture has the peak performance of 1142 GOP/s and achieves up to 9.19x and 3.91x speedup over high-end GPU NVIDIA 2080Ti and NVIDIA V100, respectively. Compared with latest accelerator for action recognition GCNs models, our design reaches 22.9x speedup and 28.93\% improvement on DSP efficiency.
\end{abstract}

\begin{IEEEkeywords}
Graph Neural Network, Action Recognition, Hybrid Pruning, Sparse Data Compress, Hardware Accelerator, Field-programmable Gate Array (FPGA)
\end{IEEEkeywords}

\section{Introduction}
Action recognition based on deep learning has great potential to be applied in kindergartens and hospitals to prevent danger motions.
Skeleton-based graph convolutional networks (GCNs) methods have achieved state-of-the-art (SOTA) prediction accuracy in the field \cite{1}\cite{2}\cite{3}.
Mature pose estimation algorithms extract human skeletons from video stream with real-time speed, for example, OpenPose \cite{4} and Alphapose \cite{5}.
GCNs action recognition models and pose estimation models thus can be combined into an end-to-end system.

Despite skeleton-based GCNs having great advantages, several problems limit their applications in expected scenarios.
Firstly, intensive computation and large network architectures are embedded in skeleton-based GCNs, causing great computing cost on GPUs.
MobilePose \cite{6} can produce human skeletons on mobile platform Snapdragon 845 with 60 fps and 44.4 fps/Watt,
while 2s-AGCN model merely has a performance of 28 fps and 0.11fps/Watt on NVIDIA 2080Ti GPU.
The computing speed and power-consumption's gap indicates a great importance on accelerating GCNs action recognition algorithms.
Secondly, the expected application environment of action recognition models poses stringent constraints on power-consumption and throughput.
However, the high-performance GPU cannot meet the power-efficiency demand.

Network pruning and graph sparsification \cite{7}\cite{8} are two effective methods to relieve model complexity.
However, these methods are unsuitable for skeleton-based GCNs. There are two reasons. (i) \emph{Dataflow is transformed}: Graph computation changes the convolution dataflow.
When being conducted on different dataflows, traditional pruning methods for CNNs only skip useless computing in convolution but may not work on graph task. (ii) \emph{Skeleton-relationship graph is unchangeable and sensitive.} Some works use pooling \cite{7} or graph sparsification \cite{8} to drop unimportant edges and points in graph.
However, the human skeleton graph cannot be modified for bones connection being unchangeable.
Particularly, there are learnable hidden information graph \cite{1}\cite{9} which lacks sparsity in some GCNs models .
The subtle elements in such graph are proved to be positively associated with prediction performance.
For instance, in 2s-AGCN model, the prediction accuracy decreases by 2.3\% without learnable matrix \cite{9}.
Although Ding et al. \cite{10} present a FPGA-based work on accelerating ST-GCN, a smaller action recognition GCNs, their work falls short on more complex models for (i) They
do not prune the target model. (ii) They choose to compress sparse static skeleton graph, while skeleton relationship matrix in some models can be dense. (iii) Only sparsity graph is
optimized for computation, while feature sparsity is not utilized. (iv) Their single-PE design cannot meet performance requirement of application scenarios. To summarize, such GCN models with dense matrixes bring new challenges for architecture.

For these reasons, efficient pruning methods together with specific accelerator designs are urgently required to accelerate GCNs action recognition workloads.
To this end, we present RFC-HyPGCN: a runtime sparse feature compress accelerator for skeleton-based GCNs action recognition model with hybrid pruning in this paper.

A hybrid GCNs' pruning method is proposed, which reduces network parameters and skips graph computation efficiently.
We reorganize dataflow by changing the multiplication order of graph workloads and spatial convolution.
With new dataflow, graph computation and spatial convolution are skipped if the corresponding parameter is pruned as zero.
As to the temporal convolution, mixed-grained pruning method is elaborately designed.
Fine-grained pruning operation can be dealt as sampling in time series, while coarse-grained pruning is decided by spatial convolution's pruned dataflow.
Experiments demonstrate that in most cases, better prediction accuracy and more balanced pruning can be possessed by our model compared with conventional methods.
Additionally, quantization and input-skipping are applied, which are common means to accelerate neural networks.

We also design an application-specific architecture. Ten convolution blocks are mapped on FPGA, which is the platform widely used for speeding up deep neural networks.
Different from previous works, in our layer-pipelined architecture, challenges are not only reflected on different kinds of sparse tasks,
but also on how to efficiently store sparse intermediate results on chip.
Common compact formats like compressed sparse column format (CSC) is not the optimal resolution due its irregular memory access and extra encoding/decoding cost.
To address these challenges, our sparse-degree-based runtime sparse feature compress method is proposed,
which splits data encoding/decoding and corresponding storage into fine-grained bank and mini-bank.
Finally, dynamic data scheduling is applied for intra process elements (PE) to decrease the utilization of DSPs.

In summary, the contributions in this paper are:
\begin{itemize}

\item We propose a hybrid pruning method on 2s-AGCN model, which contains dataflow reorganization and mix-grained pruning method.
The experiments show that our method is better than conventional pruning on computation-skipping and prediction accuracy.

\item A co-designed architecture is implemented, including runtime sparse vector compress storage and dynamic data scheduling.
The proposed online data compact format reduces the utilization of BRAM blocks as well as keeping regularity of data-access.
Dynamic data scheduling saves DSPs and raises DSPs working efficiency with tiny delay.

\item Our design is implemented on Xilinx XCKU-115 FPGA platforms with 172 MHz.
It achieves 2.61x-9.19x accelerating ratio compared with NVIDIA 2080Ti and 1.36x-3.91x with NVIDIA V100.
On contrast to similar work \cite{10}, ours exceeds on both peak performance and DSPs efficiency.
It has the potential to apply in end-to-end and low-power real time environments.
\end{itemize}
\section{Background on 2s-AGCN Model}
\begin{figure}[t]
\centerline{\includegraphics[width=0.42\textwidth]{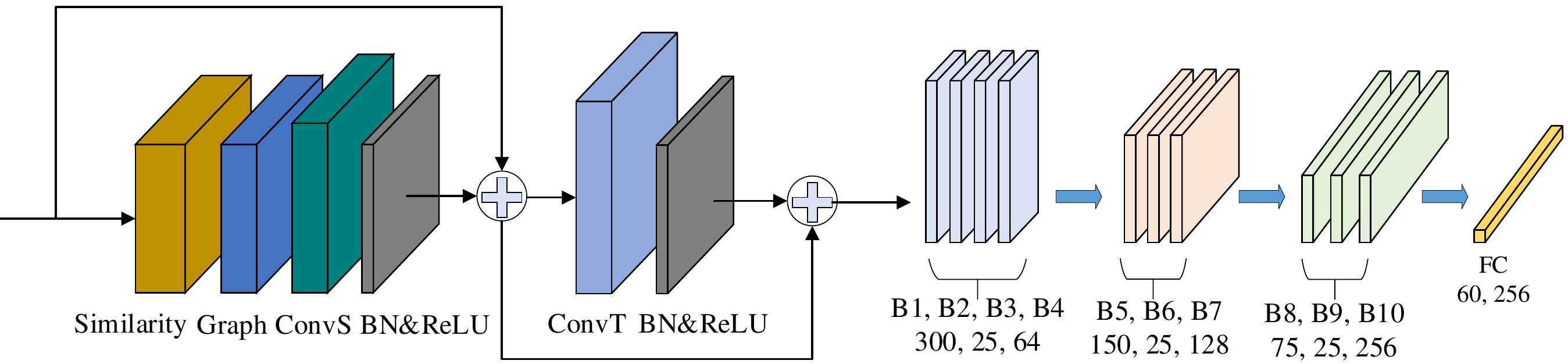}}
\caption{Left: Structure of the basic convolutional block. ConvS stands for spatial convolution and ConvT stands for temporal convolution. Right: Variance of the feature dimension. There are 25 key joints in human skeleton and 300 skeleton vectors in the original input feature.}\label{myfig1}
\end{figure}
The skeleton-based action recognition GCN models regard human skeleton features as input and predict human action , like waving and drinking.
Several human skeleton datasets have been proposed, for example NTU-RGB+D\cite{11} and Kinetics \cite{12}.
Our experiments on 2s-AGCN are trained and tested on NTU-RGB+D. There are ten convolutional blocks and one fully-connected (FC) layer in 2s-AGCN model.
As shown in the left picture in Fig.~\ref{myfig1}, the computation in each block can be divided into five phases:
graph computation, self-similarity computation, spatial convolution, temporal convolution and shortcut connection.
Batch-normalization and ReLU activation follow behind each convolution operations. With network going deeper, more channels are stacked on feature.
The right picture of Fig.~\ref{myfig1} illustrates this tendency in data dimension.

In each layer, three different graphs are involved in computation: $A_{k}$, $B_{k}$ and $C_{k}$, $k$ is explained in \eqref{eq2}. The first part $A_{k}$ is the static human skeleton graph, the second part $B_{k}$ is a learnable skeleton connection graph and $C_{k}$ is a data-dependent graph generated from self-similarity process. Elements in $B_{k}$ are trained to indicate hidden relationships between joints and bones. Unlike static graph $A_{k}$, $B_{k}$ is dense and sensitive to numerical changes. In \cite{9}, $C_{k}$ is produced via \eqref{eq1}, where high-dimension tensor transposition and multiplication are conducted on input feature. $W_{\theta}$ represents similarity coefficient. To sum up, the computation of graph and spatial convolution can be described as \eqref{eq2}. $\bigotimes$ represents convolution operation, $K_{\nu}$ denotes the neighbour size of the graph computation and is set to 3 in the 2s-AGCN model. The kernel size of spatial convolution¡¯s weight $W_{k}$  is set to 1.
\begin{equation}
C_{k}=softmax(f_{in}^{T}W_{\theta}f_{in})\label{eq1}
\end{equation}
\begin{equation}
f_{out}=\sum_{k}^{k_{\nu}}f_{in}(A_{k}+B_{k}+C_{k})\bigotimes W_{k}\label{eq2}
\end{equation}

Different from $A_{k}$ and $B_{k}$ which are determined before inference, $C_{k}$ relies on input feature, thus needs runtime computing for each prediction.
Table.~\ref{tab1} demonstrates the computing cost of self-similarity. The running performance of 2s-AGCN with and without $C_{k}$ are tested on NVIDIA V100.
 At the cost of computing complexity and longer time-delay, $C_{k}$ only elevates prediction accuracy by 0.3\%.
 From the view of software-hardware co-design, dropping $C_{k}$ graph is a reasonable trade-off for workload reduction.
\begin{table}[t]
\caption{Model performance with(w/) and without(w/o)$C_{k}$.}
\begin{center}
\begin{tabular}{|c|c|c|c|}
\hline
 &accuracy&throughput&power efficiency \\
\hline
2sAGCN(w/C)&93.70\%&69.38 fps&0.28 fsp/watt \\
\hline
2sAGCN(w/oC)&93.40\%&98.87 fps&0.40 fps/watt \\
\hline
\end{tabular}
\label{tab1}
\end{center}
\end{table}
\begin{figure}[htbp]
\centerline{\includegraphics[width=0.42\textwidth]{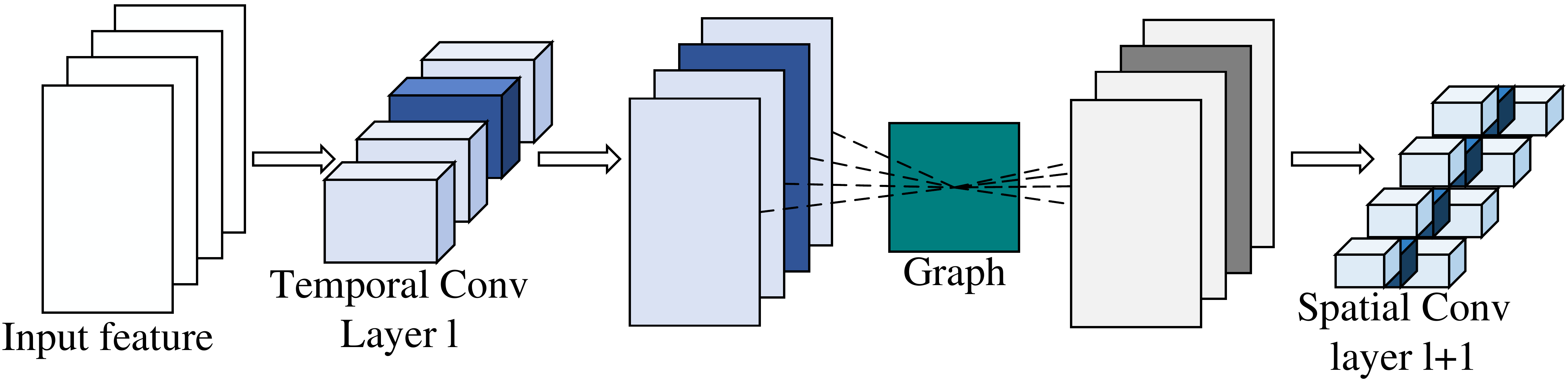}}
\caption{The neighbour connection between temporal convolutional output and spatial convolutional input.}\label{myfig2}
\end{figure}
Following the spatial convolution, temporal convolutional layer is set at the end of each convolutional block.
With kernel size of $9\times1$, temporal convolution extracts information from nine skeleton vectors in time order.
Despite the insertion of the graph computation,
temporal convolution layer in block $l$ can still be seen as the leading neighbour of spatial convolution layer in next block
because graph computation does not change temporal convolutional result along its output-channel dimension,
and spatial convolution operates indirectly on temporal convolution's output in block $l+1$ \cite{13}.
For above reasons, the connection shown in Fig.~\ref{myfig2} guides us to conduct coarse-grained pruning on temporal convolutional filters.

\section{Related Work}

\textbf{CNNs Accelerators on FPGA.} Works on FPGA-based acceleration of sparse CNNs can be categorized by different pruning granularity levels \cite{13}:
(i) for coarse-grained pruned models, (ii) for fine-grained pruned models, (iii) for mixed-grained pruned models.
Zhu et al. \cite{13} presented a zero-skipping dataflow for feature. Although such method raised computing efficiency, zero elements in intermediate result still occupied storage resource.
Lu et al. \cite{14} proposed a weight-oriented dataflow for fine-grained pruned CNNs with little decoding cost.
However, 2s-AGCN model differs from above convolutional workloads in that feature first goes through graph matrix multiplication.
This weight-oriented design cannot skip corresponding graph computation. Li et al. \cite{15} worked on PCONV pruning \cite{16}, a mixed-grained method.
With weight-stationary dataflow designed on FPGA, Li et al. improved the computing efficiency by 14.7\%-44\%.
However, this work still occupied storage space for huge scale of zero data like Lu et al., and its simple hardware structure could not tackle complex workloads in our task.

\textbf{GCN Accelerators on FPGA.} Many works on accelerating large graph's GCNs based on FPGA are presented in recent time.
AWB-GCN \cite{17} combined dynamic hardware configuration and runtime hardware workloads balancing on several large graph datasets.
Zhang et al. \cite{18} partitioned input data into smaller segments, then perform graph sparsification and node re-ordering for computation reduction and data locality.
To sum up, above works focus on: (i) Leveraging and expanding graph adjacency matrixes sparsity,
(ii) Avoiding irregularity and randomness of data distribution in graph computation,
(iii) Keeping balanced workloads between PEs or computing phases, via offline and online ways.
Unfortunately, graph in skeleton-based GCNs for action recognition models is dense and unchangeable.
The data sparsity is embedded in temporal feature and pruned weights, not the graph.
Moreover, action recognition GCNs behave not only like CNNs, but also like graph processing, leading to graph-specific design requirements.
Therefore, current specialized architectures on CNNs and GCNs cannot efficiently accelerate target models since they just take one of the two sides.


\section{Methodology}
This section introduces our hybrid pruning method for action recognition GCNs. The dataflow reorganization, coarse-grained and fine-grained pruning on temporal convolution are described respectively.
\subsection{Dataflow Reorganization}
After clipping self-similarity graph, the computing flow between graph and spatial convolution can be further summarized as \eqref{eq3}, where $G_{k}$ denotes $A_{k}+B_{k}$ from \eqref{eq2}. The computing order is first high-dimensional matrix multiplication with $G_{k}$, then the spatial convolution of $W_{k}$ and finally the result merging of three loops. In this dataflow, common pruning methods only functions in second phase but cannot optimize the graph computation, which occupies 49.83\% of total workloads in \eqref{eq3}.

\begin{equation}
f_{out}=\sum_{k}^{k_{\nu}}f_{in}G_{k}\bigotimes W_{k}\label{eq3}
\end{equation}

To better analyse the dataflow, we extract first two phases and its output \textit{X}. A pixel can be described as $X(h, w, oc)$, where \textit{h, w, oc} represent height, width and output-channel coordinates respectively. Then \eqref{eq4} can be deduced from \eqref{eq3} and \textit{ic} is the acronym of input channel. Under the commutative law of multiplication, therefore \eqref{eq4} is transformed into \eqref{eq5}. By reorganizing the computing order between graph phase and convolution phase, an opportunity for graph-skipping pruning is offered here. If the parameter element $W(1, 1, i, oc)$ is pruned to zero, the graph matrix multiplication in current output channel can be ignored. Further, if we set all convolutional parameters in \textit{i} input channel as zero, then all graph computation can be skipped in current loops. The dataflow reorganization is then proposed when we apply above method to three loops in \eqref{eq3}. Unlike conventional structure pruning method which drops different channels on filters, weights in specific input channels are all set as zero on every spatial filter in current convolutional blocks. In this way, not only the convolution workload is reduced, but also the graph computation is skipped.

{\setlength\abovedisplayskip{1pt}
\setlength\belowdisplayskip{1pt}
\begin{small}
\begin{equation}
X(h, w, oc)=\sum_{i=1}^{ic}(\sum_{p=1}^{25}f_{in}(h, p, i)\times G(p, w))\times W(1, 1, i, oc)\label{eq4}
\end{equation}
\end{small}
\begin{small}
\begin{equation}
X(h, w, oc)=\sum_{i=1}^{ic}(\sum_{p=1}^{25}G(p, w)\times f_{in}(h, p, i)
\times W(1, 1, i, oc))\label{eq5}
\end{equation}
\end{small}}
 Since the graph-skipping strategy has been determined by dataflow reorganization, the next step is choosing the input channel to be pruned. Like other deep neural networks (DNN), features between convolutional layers are sparse and useful elements are unevenly distributed. Based on the observation that unstructured pruning method drops weight element with relatively small absolute value, we cut off the input channels which have least averaging absolute value. In this way, data reorganization prunes spatial convolutional weight and skips both graph and convolution computation.
\subsection{Mixed-grained Pruning Method}
In dataflow reorganization method, features in specific channels are not computed because such channels are pruned.
The coarse-grained method then prunes corresponding temporal filters via connections in Fig.~\ref{myfig2} with no extra accuracy loss.
Moreover, this neighbour connection is hardware-friendly for that the number of pruned channels in spatial filters equals the number of pruned filters in temporal convolution.
This inherent feature supports a balanced layer-pipelined architecture.
\begin{figure}[t]
\centerline{\includegraphics[width=0.32\textwidth]{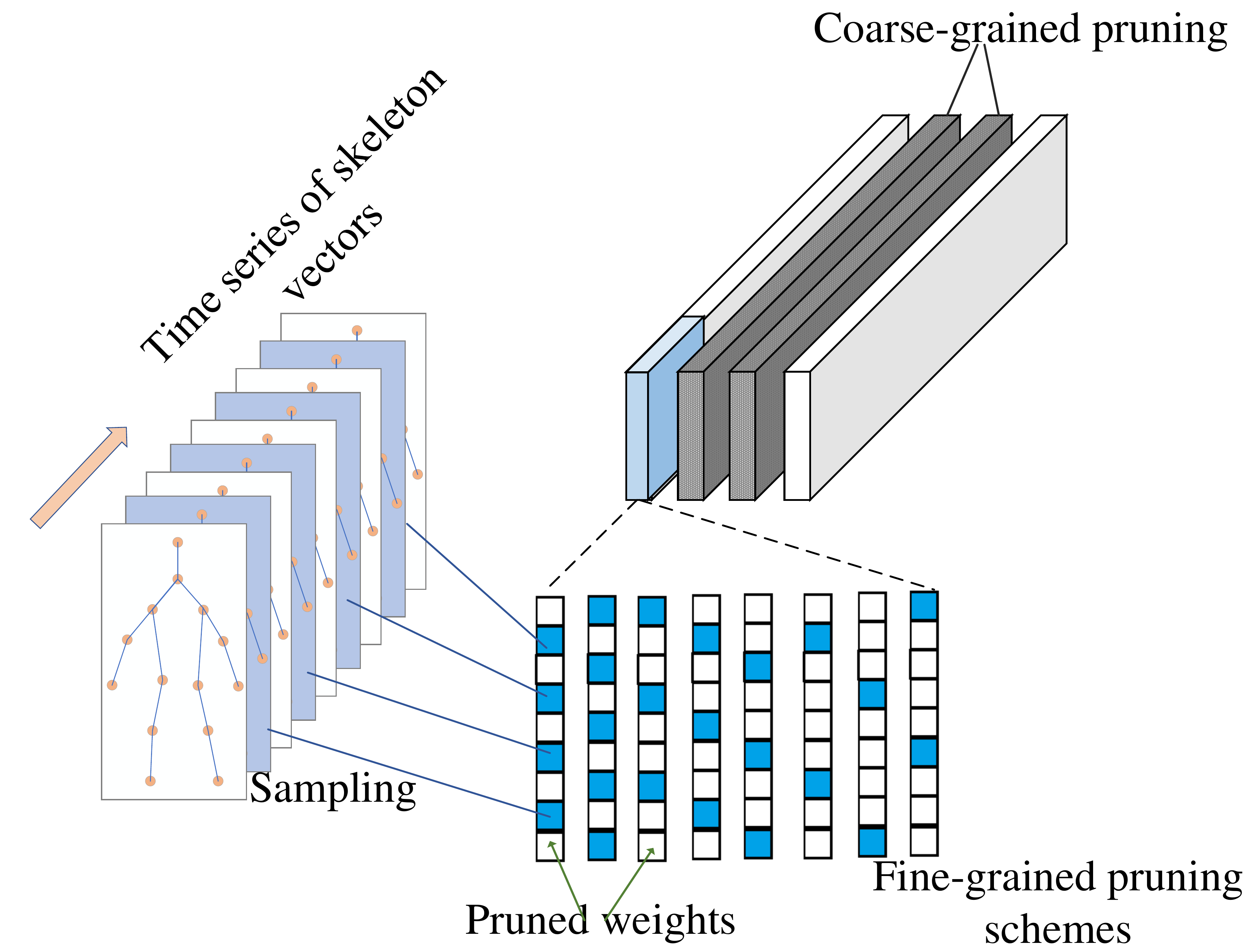}}
\caption{The illustration of fine-grained pruning on temporal convolution. White elements are pruned while blue ones are kept. Every 9$\times$1 kernel performs on time series of skeleton vectors, and the blue vectors are sampled by first pruning scheme.}\label{myfig4}
\end{figure}

Coarse-grained pruning can provide 49.83\%-88.96\% compression ratio on temporal filters, depending on the pruning scheme in data organization phase. To further prune temporal convolutional weights, fine-grained pruning is proposed. The point of fine-grained method is that in temporal convolution, zero weight means not sampling current vectors in time order. Fig.~\ref{myfig4} demonstrates details of sampling-like fine-grained pruning method. Several pruning schemes with different intervals and offsets are conducted on filters recurrently. By this means, the design of pruning scheme is turned into a sampling problem. We can simulate various sampling schemes on filters, with different sampling frequencies and phases represented by intervals and offsets. Experiments show that with proper pruning scheme, our fine-grained method can keep accuracy as well as discarding unimportant weight.

Conventional unstructured pruning methods randomly drop the weight elements with least absolute value, which are expensive and unbalanced on hardware. However, with determined cavity schemes, our fine-grained pruned model can be indicated by structured weight together with masks. Furthermore, we guarantee the balancing distribution of reserved weight by controlling start-points of different sampling patterns. Like Fig.~\ref{myfig4} shows, in a loop of eight different pruning modes, weight elements in every position of kernels are evenly kept by two or three times. Compression ratio can be adjusted via fine-grained pruning design.
\begin{figure}[t]
\centerline{\includegraphics[width=0.45\textwidth]{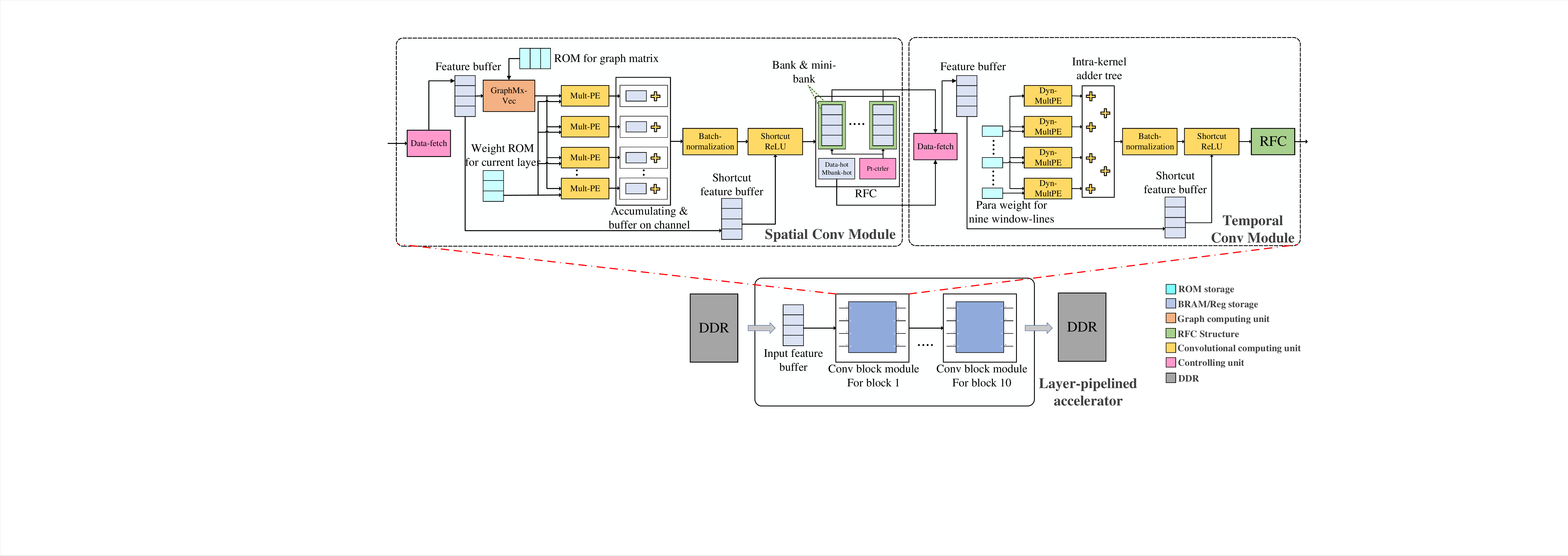}}
\caption{The demonstration of overall architecture.}\label{myfig5}
\end{figure}
\section{Architecture}

This section introduces the detailed architecture of our accelerator, where all pruned convolutional blocks are mapped on chip.

\emph{Overview}: Fig.~\ref{myfig5} depicts the overall design of our layer-pipelined architecture. Conv block module constitutes the whole architecture, containing one spatial conv module (SCM) and one temporal conv module (TCM). To be more detailed, one conv block module processes on convolution block. Due that proposed pruning method reduces model size, all parameters and graph are stored in ROM storage on chip. Mult-PE is the basic computing unit of SCM, while Dyn-Mult-PE is the basic unit in TCM. Moreover, runtime sparse feature compression module (RFC) functions at the junctions of different layers to compact and store temporal results.

\subsection{Spatial Conv Module}
The main task of SCM is performing graph computation and pruned spatial convolution. Data-fetch controls the address of data-loading and decodes compact feature into sparse form, which is prepared for computing. Feature buffer receives and stores decoded data in order. The decoding process will be explained later. Sparse feature will first multiply with graph vectors, then conduct convolution with non-zero weight in Mult-PEs. Pruned channels are skipped and multiplication results are summed up in accumulating buffer on output channel. After batch-normalization operation, dataflow merges with original input activation, which is stored in shortcut feature buffer. ReLU function is combined with encoding function parts.

\begin{figure}[t]
\centerline{\includegraphics[width=0.42\textwidth]{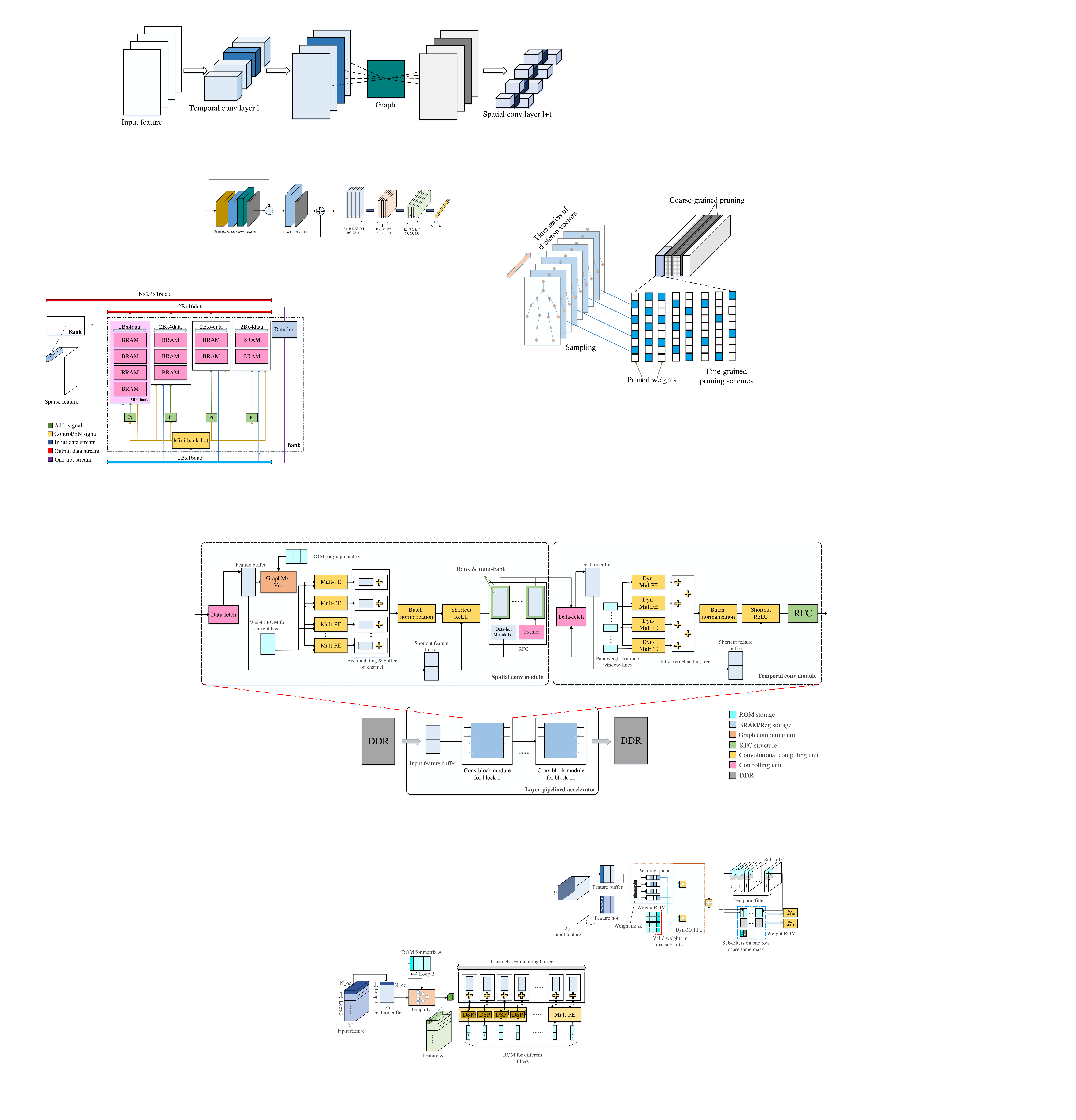}}
\caption{Illustration of SCM dataflow organization.}\label{myfig6}
\end{figure}

In order to combine graph computation and pruned convolution workloads, dataflow is organized as Fig.~\ref{myfig6} shows. A line in feature buffer caches 25 data and the depth equals the number of kept channels in filter. When computing, buffer offers one line of original feature data. After computation with one column vector of graph, there generates one valid element $X(h, w, oc)$ in \eqref{eq4}. Afterwards, feature buffer provides next cache line, which continues to produce $X(h, w, oc+1)$. Following this mode, when all output elements on current output channel are computed, feature buffer returns to the first line and graph ROM switches to the vector in next column to prepare for $X(h, w+1, 0: N_oc)$. When the workload of one row feature tensor is finished, feature buffer receives next row of tensor to start a new sub-loop. In this way, feature is produced in a channel-first order. Our dataflow reorganization method essentially abandons feature data on specific channels, so we skip corresponding workloads by not sending them to feature buffer.

Feature element is broadcast to all Mult-PEs. In the same channel-first order, weight ROM sends different filters' parameters into computing units. To match the pruned model, only non-zero weights are stored. Each Mult-PEs includes four DSPs, and by adjusting the number of Mult-PE, our design can fit into different layers. Results from parallel Mult-PEs are accumulated and buffered on channel direction as well. When the sum counter reaches the number of valid channels, current data will be transferred into post-processing modules.

\subsection{Temporal Conv Module}
\begin{figure}[t]
\centerline{\includegraphics[width=0.42\textwidth]{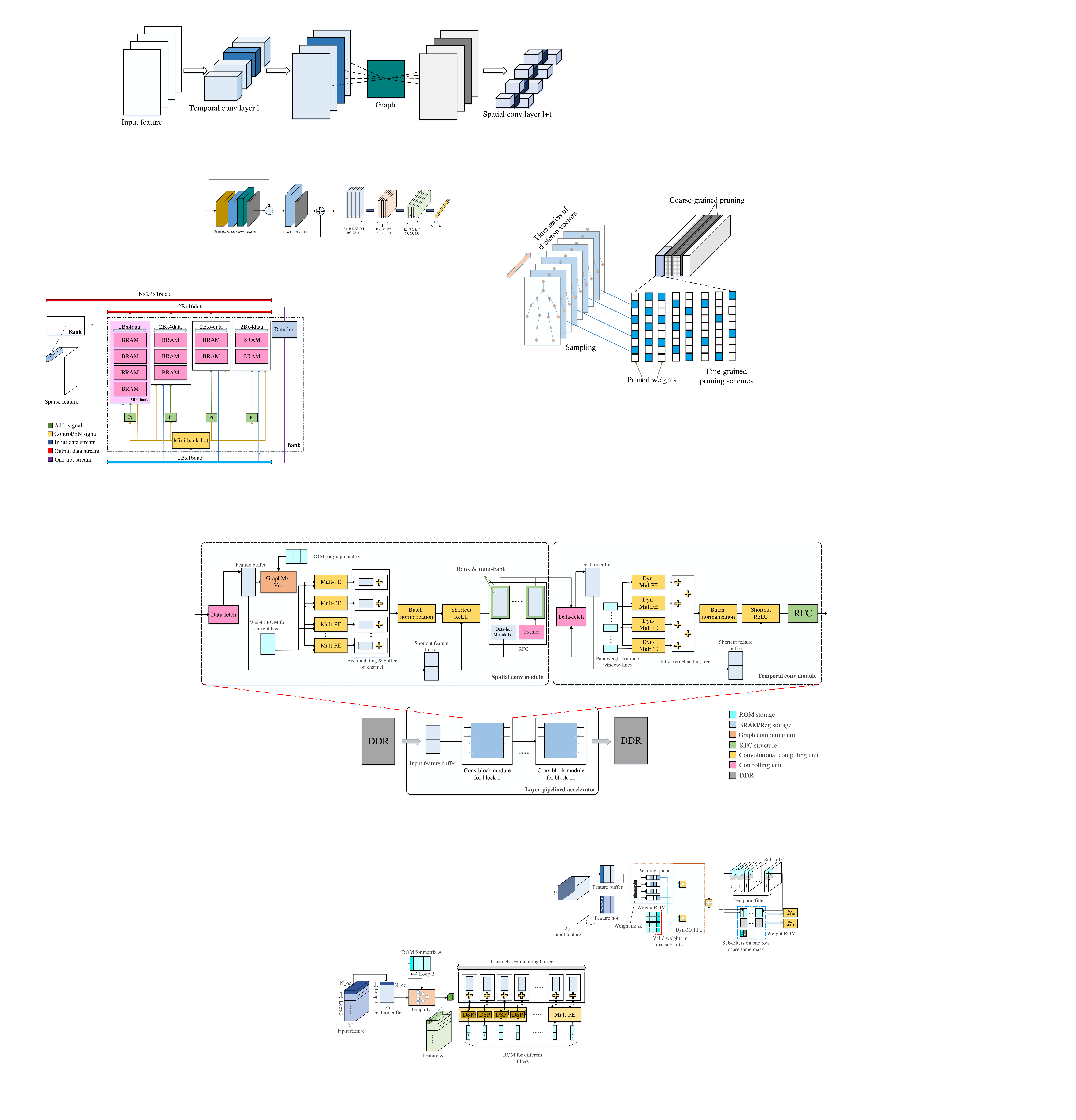}}
\caption{Illustration of TCM's dataflow.}\label{myfig7}
\end{figure}

TCM is designed to accelerate temporal convolution workloads, whose kernel size is $9\times1$. Fig.~\ref{myfig7} shows the detailed information of TCM. Similarly, feature buffer stores decoded data from data-fetch. However, buffer width is turned from 25-data into 9-data, and the depth is tuned for holding an $9\times25\times in\_c$ area of feature tensor. Additionally, feature's one-hot code is sent from data-fetch to feature hot storage as well. Valid weight together with its masks is stored on chip. As depicted in Fig.~\ref{myfig4}, several balanced fine-grained pruning schemes are conducted on leftover filters in recurrent ways, providing opportunities for structured weight storage. One temporal filter is divided into several $1\times 1\times 16$ sub-filters, thus, parameters can be folded into sub-filters format and then be stored in a recurrent mode. Moreover, Dyn-Mult-PEs are put across input channel and parallelizes on filter's rows. There are two reasons: (i) This parallel scheme can directly skip the abandoned filters in coarse-grained pruning, (ii) Each row of sub-filters is taken by one Dyn-Mult-PE and each function part handles one row of weight tensor. In this way, one Dyn-Mult-PE only needs to process weights derived from static cavity mask, such as four or six weights in Fig.~\ref{myfig7}. This design further eliminates data irregularity and scheduling uncontrollability.

Shown in Fig.~\ref{myfig7}, one row of a sub-filter is assigned to a Dyn-Mult-PE, which includes four or six waiting queues and several DSPs.
Logic and operantion is performed first on weight mask and feature mask to skip the zero-feature and dropped weights.
Then, valid feature enters one waiting queue, which is bonded to a non-zero weight in the sub-filter.
To decrease the use of DSPs, dynamic data scheduling is designed by dispatching data from busy waiting lines to empty DSPs.
Because multiplication in a Dyn-Mult-PE is part of the intra-filter computation, results need to be summed up and afterwards sent to the adder-tree.
While dynamic data scheduling has advantages on saving DSP resource, it may increase the working delay at workloads-intensive cases.
With the help of recurrent fine-grained pruning and statistic sparsity, we calculate the expectation of valid computation in one sub-filter and use it to guide the DSP occupation.
The detailed method is illustrated in \eqref{eq6}, where the number of kept weight in a sub-filter is assumed as six and $s$ stands for feature sparsity.
\begin{equation}
E(D)=\sum d\times p(d) = 3(1-s)^{3}+3s^{2}(1-s)+6s(1-s)^{2}\label{eq6}
\end{equation}
\subsection{Runtime Sparse Feature Compress}
\begin{figure}[t]
\centerline{\includegraphics[width=0.42\textwidth]{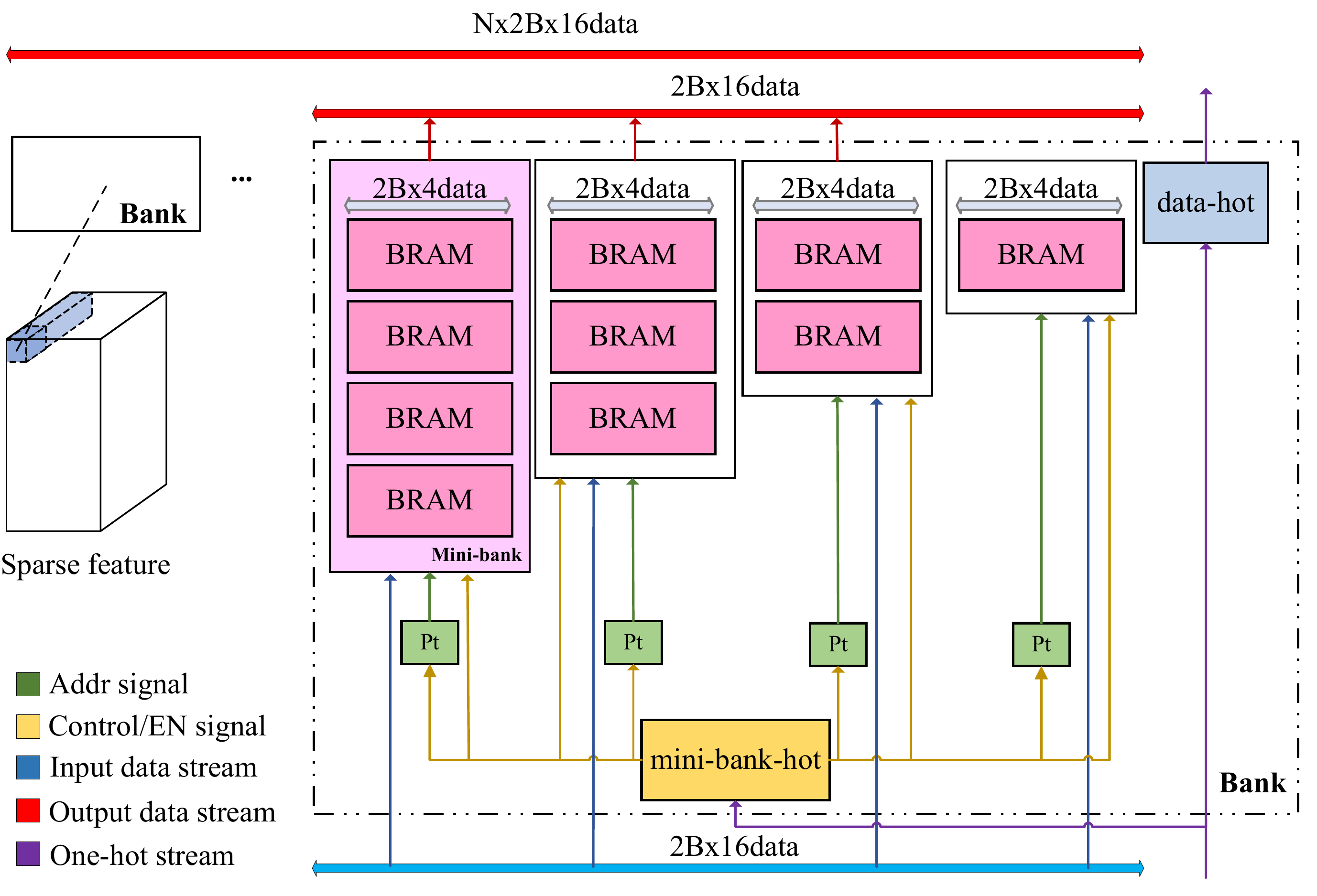}}
\caption{Illustration of TCM's dataflow.}\label{myfig8}
\end{figure}

Despite layer-pipelined architecture poses great advantages on throughput, it has to store massive temporal computing results for shortcut task. RFC is presented to address this issue. Encoding, compact storage and decoding are included in RFC structure. Encoding process is combined with ReLU while decoding is embedded in data-fetch module. The whole structure of RFC is displayed in Fig.~\ref{myfig8}.

\textbf{Encoding:} At first, one feature vector is divided into several banks across channels. The width of each bank is 16 data-wise. ReLU function parts perform on banks, providing activation and one 16-bit hot code, which denotes the positive/zero value. Then valid elements are gathered at higher bits while unused bits are padded with zero. After that, mini-bank-hot code (mbhot) is generated according to the number of non-zero data in bank. Mbhot indicates which mini-banks are used in the bank storage. Encoding parts work in pipeline and during several working cycles, the whole vector is finally turned into compact format. Instead of compressing one vector as whole, we lower the encoding cost by setting bank as the finest grain of ReLU and encoding process.

\textbf{Storage:} The compact storage consist of bank storage units, which includes mini-banks, mini-bank-hot storage, data-hot storage and address controller pt. The key insight of bank storage design is to keep access regularity on data-width dimension and reduce useless storage on data-depth dimension. When input data and hot codes are valid, mbhot sends enabling signal to every mini-banks and related pts. For example, if the input data-hot code is 0001\_1100\_0000\_0111, meaning there are five non-zero data in the high bits of input data stream and the mbhot is 1100. The first mini-bank receives and stores four valid data, the second mini-bank keeps fifth valid data and three zero data. Their pt will self-add after this data-writing. Other mini-banks and their pt are not started. Similarly, when we need to load data from bank storage in order, mbhot enables related mini-banks and pts to output correct data. The output of disabled mini-banks is covered by zero. Without any random access schemes, compact data can be both stored and loaded in only one cycle.

Another issue on compact storage is to determine the volume of each mini-bank. Like deciding DSP's utilization, we can calculate the expectation of useful data based on offline sparsity. However, there always exists vectors drifting away from average sparsity. Denser vectors demand deeper mini-banks on the tail (the rightmost mini-bank in Fig.~\ref{myfig8}) while the lighter vectors merely occupy head mini-banks. In ideal cases, every vector is fit in bank-lines with no mini-banks unused and no vector truncated, but it is hard to precisely determine the number of valid data in every vector. Feature¡¯s sparsity distribution of each layer can help us to adjust the depth of mini-banks. For example, sparsity of feature is 50\% and 25\% of vectors has sparsity higher than 75\%, 25\%'s feature¡¯s are between 50\%-75\%, 25\%'s data is between 25\%-50\% and 25\% vectors' are below 25\%. The mini-bank arrangement in Fig.~\ref{myfig8} meets the demand of different density feature and reduces 37.50\% storage resource compared with sparse format. In actual design, BRAM units have variable grains, which provides more flexibility.

\textbf{Decoding:} The decoding function is integrated in data-fetch module in SCM and TCM. Data-fetch not only controls loading address, but also translates compact data into sparse form. After receiving both data stream and data-hot codes from bank storages, parallel decoding modules perform on each banks¡¯ output. Each translation part processes compressed feature in four pipeline stages, four data for one stage. Matched with encoding phase, the output of one bank is seen as the basic decoding grain, which further decreases the complexity of decoding circuit.

\section{Experiments}
In this section, we evaluate our design on both software and hardware views.
Our pruning method is explored on one V100 GPU using PyTorch, and accelerator architecture is implemented with Verilog HDL on Vivado 2018.3 IDE.

\subsection{Validations on Hybrid Pruning Method}
In experiments, the proposed hybrid pruning method on 2s-AGCN model is compared with unstructured pruning on NTU-RGB+D,
which contains 37k training and 18k testing data.
We explore the impact of different pruning designs on accuracy,
e.g.,  various fine-grained pruning schemes for temporal filters and channel-dropping modes in reorganization phase.

\begin{figure}[b]
\centerline{\includegraphics[width=5.5cm,height=4cm]{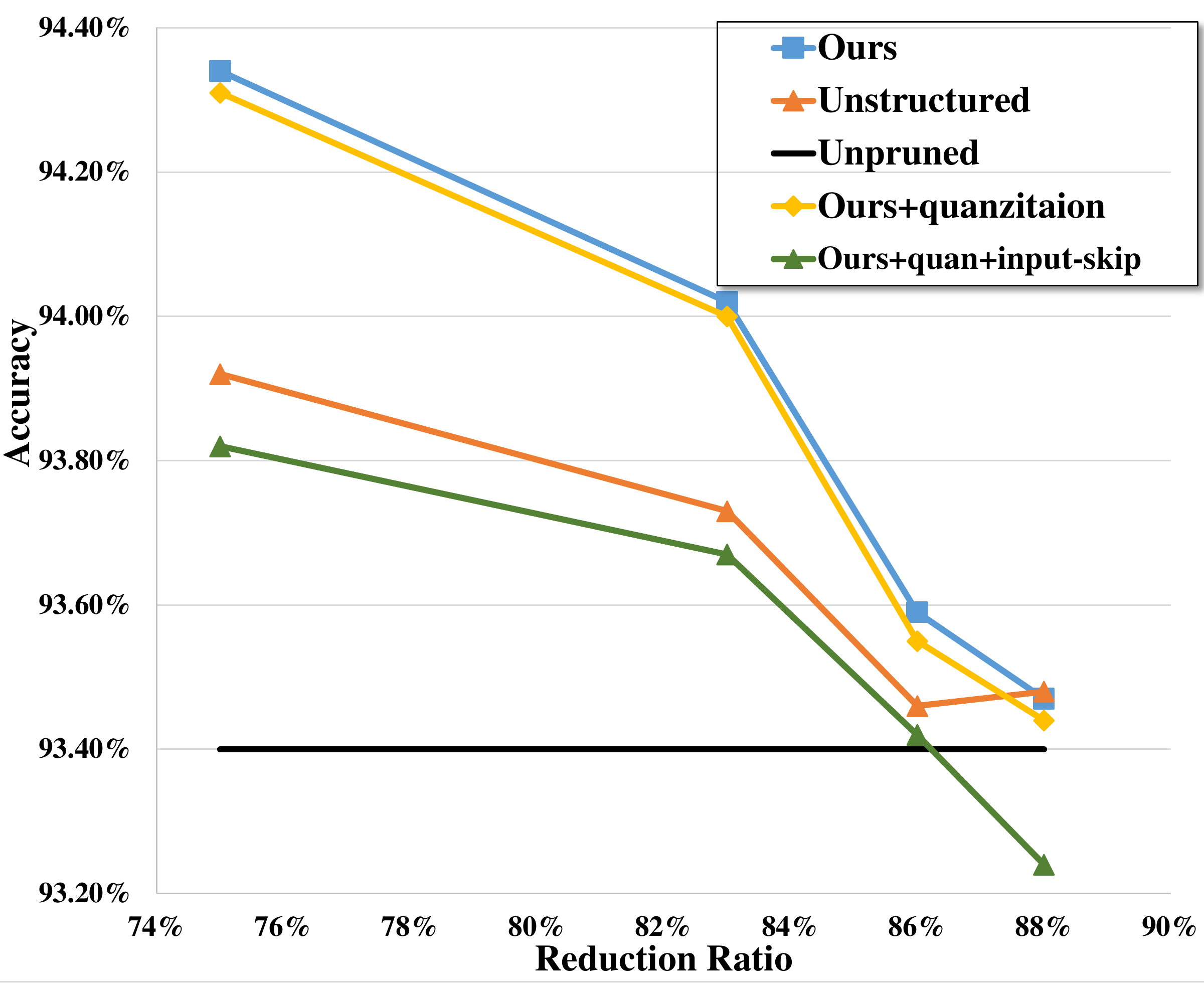}}
\caption{Comparison results with unstructured pruning.}\label{myfig9}
\end{figure}

\textit{Comparison:} Fig.~\ref{myfig9} illustrates the contrast between our hybrid pruning methods and conventional pruning means.
Both unstructured pruning and hybrid pruning can elevate accuracy for deleting some unimportant weights and improving convergence performance.
With same parameters reduction rate, our method achieves better accuracy performance in most cases. Additionally, we apply quantization on our pruned models.
With negligible accuracy loss, float data is transformed into fix-point format, where eight bits are allocated to decimal part and eight to integer part.
To further accelerate the proposed application-specific system, half of input skeleton vectors are skipped.
Although input-skip method lowers prediction accuracy, it brings 50\% reduction on total computation.
Besides, the input-skip model with 86\% compress ratio still keeps the accuracy no less than original model, so we choose this model as final accelerating target.

\begin{figure}[t]
\centerline{\includegraphics[width=8.5cm,height=3.5cm]{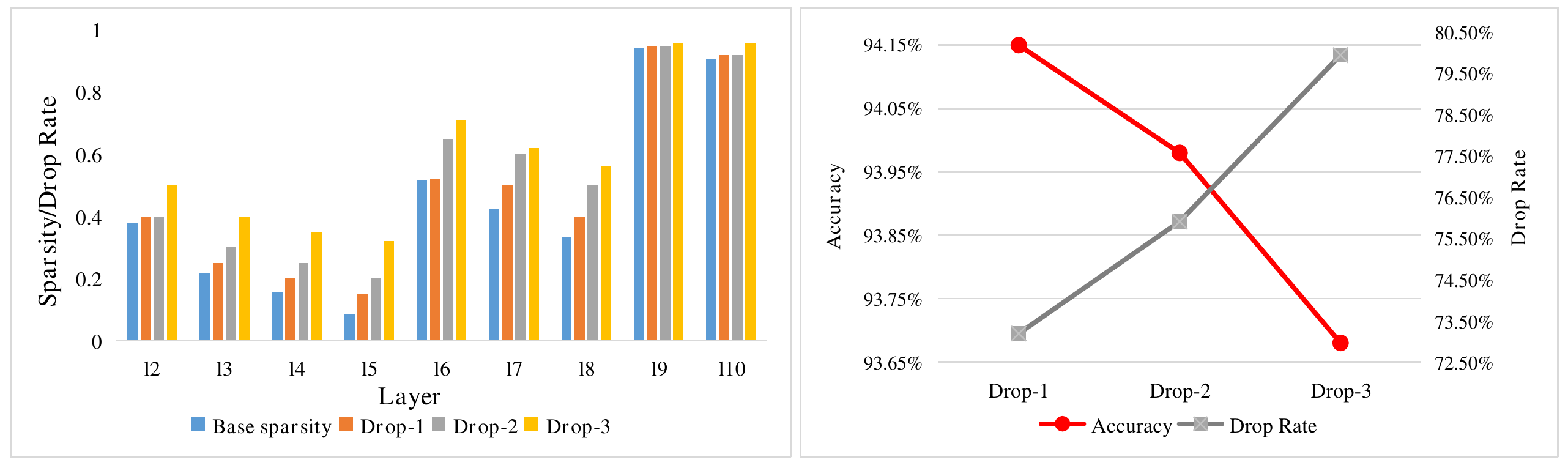}}
\caption{Exploration on channel dropping.}\label{myfig10}
\end{figure}

\textit{Exploration:} Both data reorganization and fine-grained temporal pruning are fatal to model accuracy.
To find the best pruning scheme, we conduct isolated experiments respectively. Additionally, based on our dataflow reorganization method, graph-skipping rate equals channel-dropping rate of this phase.
 Guided by feature sparsity in Fig.~\ref{myfig10},
we first set each layer¡¯s channel pruning rate roughly equals it¡¯s sparsity respectively. For higher parameter reduction ratio,
we progressively raise compressing rate on layers and observe effect on accuracy, as shown in Fig.~\ref{myfig10}. Drop-1, Drop-2, and Drop-3 stands for different channel pruning design.
The spatial convolutional parameter in block 1 is not pruned for it only has three input channels.
Also, mix-grained pruning on temporal convolution is excluded to validate data reorganization method.
It reveals that with compress rate shifting away from base sparsity, model reduction is growing while accuracy is decreasing.
We choose the Drop-1, which keeps the best accuracy, as our base dataflow reorganization design.

\begin{figure}[b]
\centerline{\includegraphics[width=0.42\textwidth]{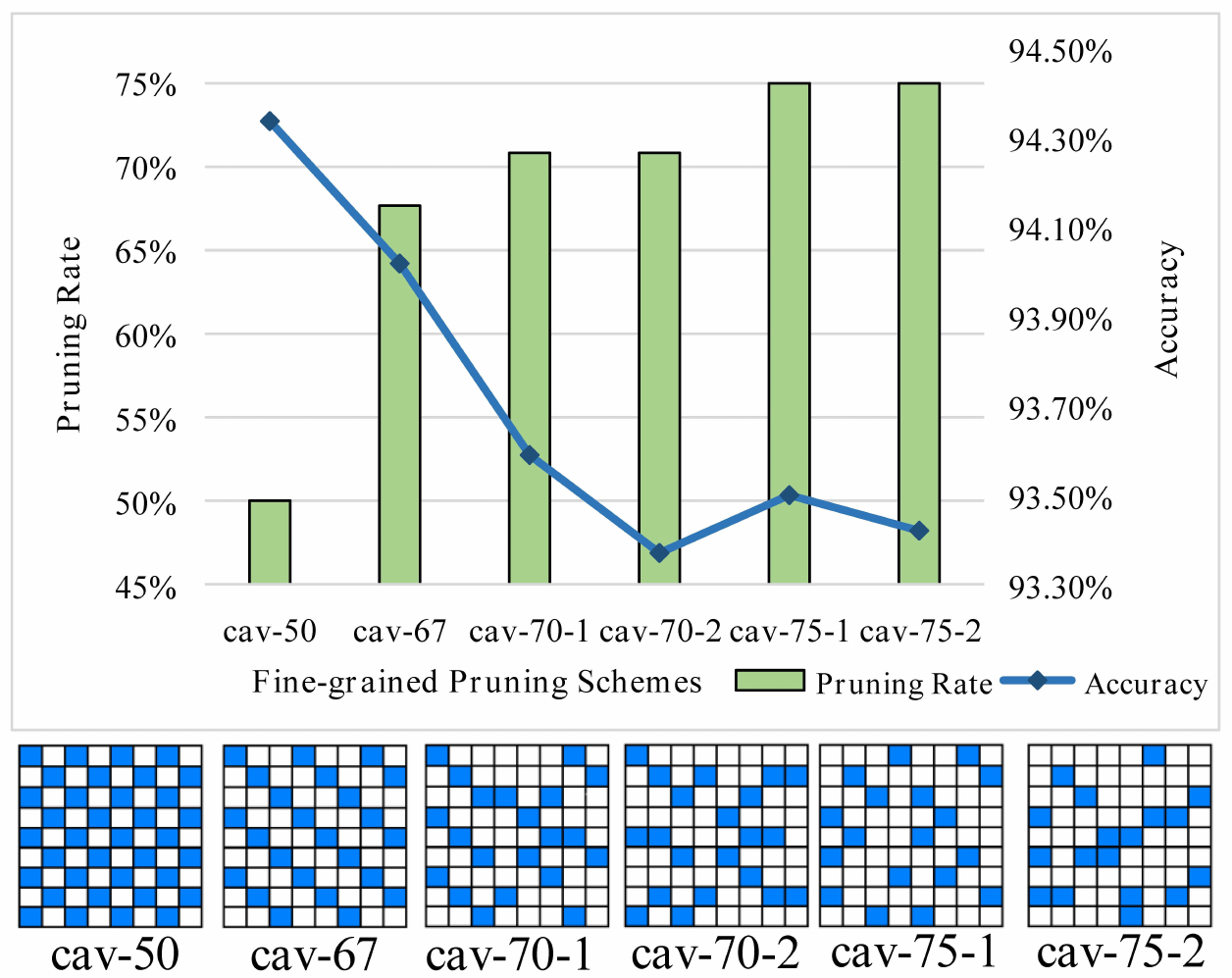}}
\caption{Exploration on fine-grained pruning schemes.}\label{myfig11}
\end{figure}

The fine-grained method is important in holding accuracy and keeping balanced-pruning, since coarse-grained means is totally decided by data reorganization.
We carry out several experiments on fine-grained pruning, including different pruning intervals, offsets and pruning rates.
All experiments are based on Drop-1 model in Fig.~\ref{myfig10} and results are shown in Fig.~\ref{myfig11}.
Pruning schemes in Fig.~\ref{myfig11} are named as the combination of cav (cavity), pruning percent (50, 67 for instance) and intra-order.
The size of rectangles below is $9\times$8, which denotes eight $9\times$1 kernels in loop.
Cav-70-1 means the first cavity patterns with 70\% reduction rate. With reduction rates expanding, model bears more accuracy loss in general.
However, cavity patterns play an important role as well.
With same compress ratio of 70\%, cav-70-1 performs better than cav-70-2 on accuracy for more balanced weight pruning.
Every weight line in cav-70-1 has two or three sampling chances, while in cav-70-2, different lines are kept from one time to four times.
Balanced pruning schemes not only provide convenience for hardware, but also ensure the accuracy performance.
The same situation happens between cav-75-1 and cav-75-2 as well.
 Taking both compress ratio and accuracy into consideration, cav-70-1 is chosen to be the final design.

\subsection{Hardware implement}
\textit{Dyn-MultPE:} Dyn-MultPE works on the cav-70-1 cavity pattern, which means there are three Dyn-MultPEs needing to six waiting queues and six facing four waiting queues.
 Based on \eqref{eq6}, different numbers of DSPs are settled in each layer¡¯s Dyn-MultPEs.
 We also adjust the number of temporal convolutional PE to keep balance between pipeline stages.
 We choose some detail information to show in Table.~\ref{tab2}, where our dynamic data scheduling trades only 6.48\% of longer delay for DSP reduction of 23.24\%.

\begin{table}[t]
\caption{Utilization, working efficiency and max delay of Dyn-MultPE.}
\begin{center}
\begin{tabular}{|c|c|c|c|c|}
\hline
layer&DSP in one PE&total DSP&efficiency &max delay\\
\hline
1&4/6&63&66.79\%&0.00\%\\
\hline
2&4/6&126&83.76\%&3.70\%\\
\hline
3&4/6&126&80.96\%&0.00\%\\
\hline
4&2/3&126&83.46\%&7.40\%\\
\hline
total&&882&75.38\%&6.48\%\\
\hline
static&&1149&57.86\%&0.00\%\\
\hline
\end{tabular}
\label{tab2}
\end{center}
\end{table}

\begin{table}[t]
\caption{Feature sparsity distribution of some layers}
\begin{center}
\begin{tabular}{|c|c|c|c|c|}
\hline
layer&I&II&III&IV\\
\hline
l1.sconv&$<$0.01\%&29.35\%&70.64\%&$<$0.01\%\\
\hline
l1.tconv&0.02\%&94.73\%&5.25\%&0.00\%\\
\hline
l2.sconv&0.00\%&0.73\%&75.79\%&23.48\%\\
\hline
l2.tconv&$<$0.01\%&34.24\%&65.76\%&0.00\%\\
\hline
\end{tabular}
\label{tab3}
\end{center}
\end{table}

\textit{RFC:} As stated above, RFC design relies on sparsity distribution.
To optimize runtime compress storage, we refer to pruned model's offline sparsity distribution, as is partly shown in Table.~\ref{tab3}.
 Feature vectors are divided into four categories by their sparsity: 75\%-100\% (I), 50\%-75\% (II), 25\%-50\% (III) and 0\%-25\% (IV).
 According to our RFC design, vector of first category occupies one mini-bank, ones in II takes two, III takes three and IV takes four mini-banks.
 We can thus get the total BRAM blocks used for RFC structure. Comparison in Fig. 11 indicates that our RFC design brings 35.93\% reduction on occupied BRAM blocks.
 Moreover, with almost same amount of used BRAM elements, RFC can finish data-loading in one cycle and encoding/decoding in four cycles,
 while CSC format usually needs 64 cycles to load data or decoding data serially. With less extra hardware cost and similar storage compress ratio,
 RFC structure achieves more regular data-access.
\begin{table}[b]
\caption{Utilization \& performance comparison between ours and \cite{10}.}
\begin{center}
\scalebox{0.70}{
\begin{tabular}{|c|c|c|c|c|c|c|c|}
\hline
&dsp&bram blocks&LUT&dsp efficiency &peak perf &frequency & fps\\
\hline
ours&3544&1806&176776&0.322GOP/s/DSP &1142GOP/S&172Mhz&271.25\\
\hline
\cite{10}&228&151&44457&0.202GOP/s/DSP &46GOP/S&188Mhz&11.99\\
\hline
\end{tabular}}
\label{tab4}
\end{center}
\end{table}
\begin{table}[t]
\caption{Performance comparison between ours and high-end GPUs.}
\begin{center}
\scalebox{0.60}{
\begin{tabular}{|c|c|c|c|c|c|c|c|}
\hline
&ours&2080Ti-original&V100-original&2080Ti(w/o C)&V100(w/o C)&2080Ti-skip&V100-skip\\
\hline
throughput&271.25&29.53&69.38&45.42&98.87&104&199.09\\
\hline
speed-up&&9.19&3.91&5.97&2.74&2.61&1.36\\
\hline
\end{tabular}}
\label{tab5}
\end{center}
\end{table}

\begin{figure}[t]
\centerline{\includegraphics[width=0.45\textwidth]{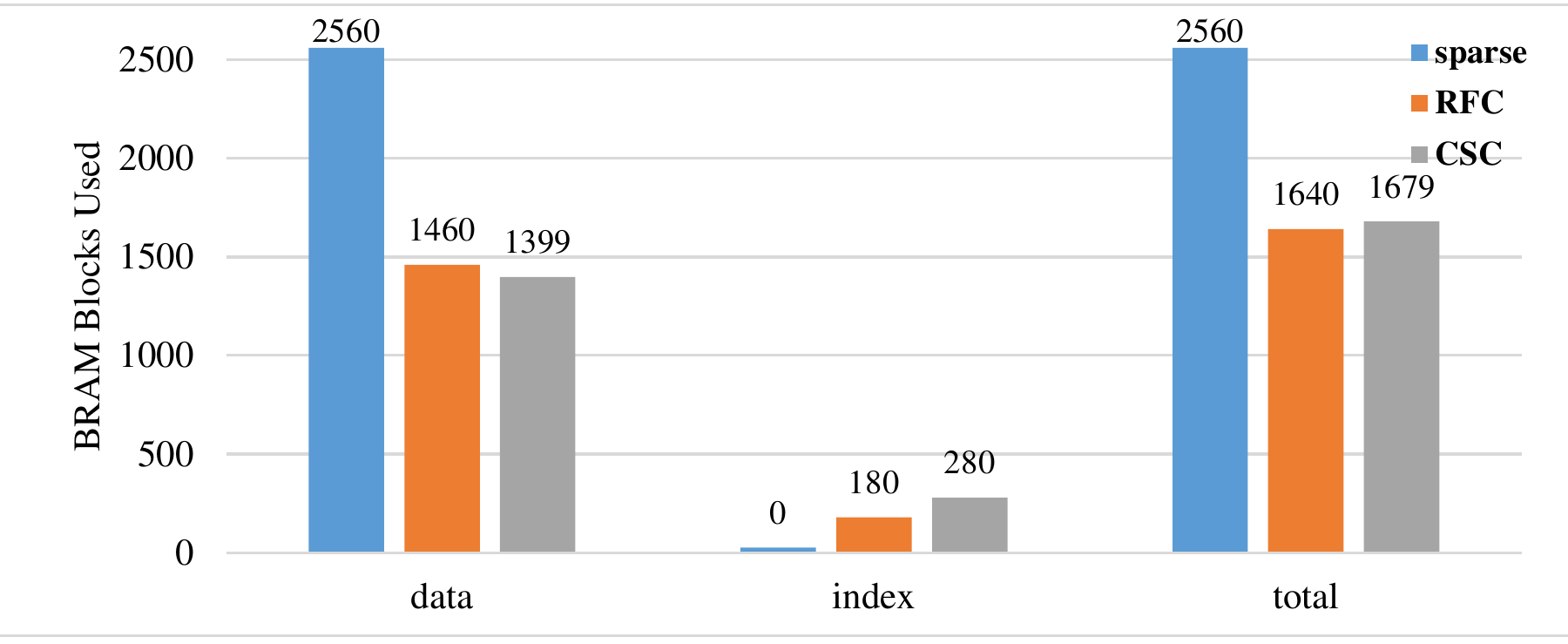}}
\caption{Storage cost of three data formats.}\label{myfig12}
\end{figure}
\textit{Overall performance}: Our architecture is implemented on Xilinx XCKU-115 with frequency of 172MHz.
The resource utilization is demonstrated in Table.~\ref{tab4}, together with comparison with \cite{10}.
Experiments have proved that our design has superiority on peak performance, throughput and DSP efficiency.
In Table.~\ref{tab5}, we compare the peak performance of ours and two GPUs.
The original means testing program is the original version of 2s-AGCN, and the skip means w/oC together with input-skipping is applied.
To fully use the memory in GPUs, target model runs with 200 or 700 samples in one batch on 2080Ti and V100, respectively.
Although V100 has 14TFLOPS of peak performance, our pruning methods provides 86\% parameter reductions and 88\% computation skipping efficiency.
Moreover, our layer-pipelined structure does not need to exchange data with DRAM except for loading original input, which further boosts performance.
Compared with two main-stream GPUs, our accelerator provides 1.36x-9.19x of speedup, showing competitive performance.
\section{Conclusion}
In this article, we propose a software-hardware co-design for action recognition GCNs: RFC-HyPGCN, including hybrid pruning method and a runtime sparse feature compress architecture. Firstly, a hybrid pruning method is explored on 2s-AGCN. Secondly, we propose an architecture based on the balanced pruned model. Finally, a runtime sparse feature compact format is designed to reduce zero-storage between layers. Experiments demonstrate that compared with conventional unstructured pruning, our method achieves better accuracy performance in most cases. The accelerator is implemented on Xilinx XCKU-115 FPGA. At the cost of negligible working delay, RFC reduces 35.93\% of used BRAM and 23.24\% of DSPs. Ours provides 22.62x speed-up and 59.41\% elevation on DSP efficiency over another work on accelerating action recognition GCNs. On contrast to high-end GPUs, RFC-HyPGCN achieves 1.36x-9.47x speed-up on throughput.

\section*{Acknowledgment}

The paper is supported by National Natural Science Foundation of China (No.61732018, No.61802420) and Pre-research Project of China (No.31513010602-1).

\bibliographystyle{ieeetr}
\bibliography{inerbib}

\begin{thebibliography}{10}

\bibitem{1}
F.~M. Bianchi, D.~Grattarola, and L.~Livi, ``Hierarchical representation
  learning in graph neural networks with node decimation pooling,'' {\em IEEE
  Transactions on Neural Networks and Learning Systems}, 2020.

\bibitem{2}
K.~Cheng, Y.~Zhang, and X.~He, ``Skeleton-based action recognition with shift
  graph convolutional network,'' in {\em IEEE/CVF Conference on Computer Vision
  and Pattern Recognition (CVPR)}, June 2020.

\bibitem{3}
J.~Gao, T.~Zhang, and C.~Xu, ``I know the relationships: Zero-shot action
  recognition via two-stream graph convolutional networks and knowledge
  graphs,'' {\em Proceedings of the AAAI Conference on Artificial
  Intelligence}, vol.~33, pp.~8303--8311, Jul. 2019.

\bibitem{4}
Z.~Cao and T.~Simon, ``Realtime multi-person 2d pose estimation using part
  affinity fields,'' in {\em Proceedings of the IEEE Conference on Computer
  Vision and Pattern Recognition (CVPR)}, July 2017.

\bibitem{5}
Y.~Xiu, J.~Li, and H.~Wang, ``{Pose Flow}: Efficient online pose tracking,'' in
  {\em BMVC}, 2018.

\bibitem{6}
https://github.com/LiQiang0307/MobilePose pytorch

\bibitem{7}
F.~M. {Bianchi} and D.~{Grattarola}, ``Hierarchical representation learning in
  graph neural networks with node decimation pooling,'' {\em IEEE Transactions
  on Neural Networks and Learning Systems}, pp.~1--13, 2020.

\bibitem{8}
J.~Li and T.~Zhang, ``Sgcn: A graph sparsifier based on graph convolutional
  networks,'' in {\em Advances in Knowledge Discovery and Data Mining},
  pp.~275--287, 2020.

\bibitem{9}
L.~Shi, Y.~Zhang, {\em et~al.}, ``Two-stream adaptive graph convolutional
  networks for skeleton-based action recognition,'' in {\em IEEE/CVF Conference
  on Computer Vision and Pattern Recognition}, pp.~12026--12035, 2019.

\bibitem{10}
L.~Ding, Z.~Huang, and G.~Chen, ``An fpga implementation of gcn with sparse
  adjacency matrix,'' in {\em 2019 IEEE 13th International Conference on ASIC
  (ASICON)}, pp.~1--4, IEEE, 2019.

\bibitem{11}
A.~Shahroudy, J.~Liu, T.-T. Ng, and G.~Wang, ``Ntu rgb+d: A large scale dataset
  for 3d human activity analysis,'' in {\em IEEE Conference on Computer Vision
  and Pattern Recognition (CVPR)}, June 2016.

\bibitem{12}
A.~Zisserman, J.~Carreira, K.~Simonyan, W.~Kay, B.~Zhang, C.~Hillier,
  S.~Vijayanarasimhan, F.~Viola, T.~Green, T.~Back, P.~Natsev, and M.~Suleyman,
  ``The kinetics human action video dataset,'' 2017.

\bibitem{13}
C.~Zhu, K.~Huang, S.~Yang, Z.~Zhu, H.~Zhang, and H.~Shen, ``An efficient
  hardware accelerator for structured sparse convolutional neural networks on
  fpgas,'' {\em IEEE Transactions on Very Large Scale Integration (VLSI)
  Systems}, vol.~28, no.~9, pp.~1953--1965, 2020.

\bibitem{14}
L.~Lu, J.~Xie, R.~Huang, J.~Zhang, and W.~Lin, ``An efficient hardware
  accelerator for sparse convolutional neural networks on fpgas,'' in {\em 2019
  IEEE 27th Annual International Symposium on Field-Programmable Custom
  Computing Machines (FCCM)}, pp.~17--25, IEEE, 2019.

\bibitem{15}
N.~Li, L.~Liu, S.~Wei, and S.~Yin, ``A high-performance inference accelerator
  exploiting patterned sparsity in cnns,'' in {\em 2020 IEEE 28th Annual
  International Symposium on Field-Programmable Custom Computing Machines
  (FCCM)}, pp.~243--243, IEEE, 2020.

\bibitem{16}
X.~Ma, F.-M. Guo, W.~Niu, X.~Lin, J.~Tang, K.~Ma, B.~Ren, and Y.~Wang, ``Pconv:
  The missing but desirable sparsity in dnn weight pruning for real-time
  execution on mobile devices,'' in {\em Proceedings of the AAAI Conference on
  Artificial Intelligence}, vol.~34, pp.~5117--5124, 2020.

\bibitem{17}
T.~Geng, A.~Li, R.~Shi, {\em et~al.}, ``Awb-gcn: A graph convolutional network
  accelerator with runtime workload rebalancing,'' in {\em 2020 53rd Annual
  IEEE/ACM International Symposium on Microarchitecture (MICRO)}, pp.~922--936,
  IEEE, 2020.

\bibitem{18}
B.~Zhang, H.~Zeng, and V.~Prasanna, ``Hardware acceleration of large scale gcn
  inference,'' in {\em 2020 IEEE 31st International Conference on
  Application-specific Systems, Architectures and Processors (ASAP)},
  pp.~61--68, IEEE, 2020.

\end{thebibliography}
\vspace{12pt}
\end{document}